\documentclass[journal]{IEEEtran}

\usepackage{amssymb}
\usepackage{amsmath}
\usepackage{amsfonts}
\usepackage{graphicx}
\usepackage{subfigure}
\usepackage{cite}
\usepackage{enumerate}
\usepackage{url}
\usepackage[noend]{algpseudocode}
\usepackage{algorithmicx,algorithm}
\usepackage{mathtools}
\allowdisplaybreaks[4]
\usepackage{epstopdf}

\begin{document}
\title{TCM-NOMA: Joint Multi-user Codeword Design and Detection in Trellis Coded Modulation based NOMA for Beyond 5G}
\author{
	\IEEEauthorblockN{
		\normalsize{Boya Di}\IEEEauthorrefmark{1}, \emph{Student Member, IEEE},
		\normalsize{Lingyang Song}\IEEEauthorrefmark{1}, \emph{Fellow, IEEE},
		\normalsize{Yonghui Li}\IEEEauthorrefmark{2}, \emph{Fellow, IEEE},
		\normalsize{Geoffrey Ye Li}\IEEEauthorrefmark{3}, \emph{Fellow, IEEE}\\}
	\thanks{The }
	\thanks{Part of this paper was presented in the International Conference on Communications (ICC)~\cite{BLYS-2018}, 2018.}
	\thanks{B. Di and L. Song are with the School of Electrical Engineering and Computer Science, Peking University, China. Y. Li is with the School of Electrical and Information Engineering, The University of Sydney, Australia. G.Y. Li is with School of Electrical and Computer Engineering, Georgia Institute of Technology, USA.}
}
\maketitle

\begin{abstract}
In this paper, we propose a novel non-orthogonal multiple access (NOMA) scheme based on trellis-coded modulation (TCM). Different from those in the traditional code-domain NOMA, the incoming data streams of multiple users are jointly coded and mapped to the codewords so as to improve the coding gain of the system without any bandwidth extension. Both codeword design and multi-user detection are considered to optimize the TCM for NOMA. The multi-dimensional (MD) constellation is carefully selected and a new set partitioning algorithm is developed for MD signal set labeling. To achieve the trade-off between the BER performance and complexity, a suboptimal two-layer Viterbi algorithm is proposed for joint decoding. Simulation results show that our proposed TCM-based NOMA scheme performs significantly better than the traditional code-domain NOMA and OFDMA schemes in terms of the BER performance under the same channel conditions.
\end{abstract}


\section{Introduction}

The past few years have witnessed the explosive growth of demand for massive data transmission brought by the increasing number of mobile devices and applications~\cite{ZLV-2016}. The traditional orthogonal frequency division multiple access (OFDMA) techniques in the fourth-generation (4G) networks fail to achieve the adequate exertion of the frequency resources due to its orthogonal nature~\cite{BLYG-2017,LYZH-2017}. This also causes access congestion especially in a dense network, urging for more efficient solutions. Capable of supporting massive connectivity and improving the spectral efficiency~\cite{ZPH-2016}, non-orthogonal multiple access (NOMA) has drawn widespread attention as a promising candidate to solve the above issues in future communications~\cite{ZXGRJV-2017,LBYSCZ-2015}.

In NOMA systems~\cite{SKBNLH-2013,DSL-201611}, multiple users are allowed to share the same resources, such as a subchannel, a time slot and so on. One of the efficient schemes is the code-domain NOMA~\cite{HH-2013,MHAH-2014}. Specifically, the incoming data of each layer are mapped to a sparse codeword in which the non-zero elements correspond to the occupying subcarriers of each layer. After superimposing the codewords of multiple layers, the transmitter sends multiple data streams to the receiver simultaneously over a limited number of subcarriers, achieving the overloading function~\cite{MZV-2019}. Each layer is assigned a pre-defined codebook where the mapping between non-zero elements and the data streams is constructed based on multi-dimensional (MD) modulation techniques~\cite{MR-2012,MHAH-2014}. To decode the superimposed codeword, the belief probability based massage passing algorithm (MPA)~\cite{YSY-2015} has been applied at the receiver.

For the code-domain NOMA, previous works~\cite{LXPD-2015,CGN-2017,DPXYD-2016,YBZJPZ-2016} have discussed various schemes to improve the coding gain. In~\cite{LXPD-2015}, a constellation optimization method has been developed in which the structure of the MD complex constellation has been optimized based on the star-QAM constellation. In~\cite{CGN-2017}, the Turbo TCM technique has been utilized to generate the multi-dimensional codebooks for each user separately, and the pure MPA has been adopted for decoding. In~\cite{DPXYD-2016}, a new constellation has been proposed and analyzed based on a novel performance criterion. In~\cite{YBZJPZ-2016}, a low-complexity detector has been designed for an uplink NOMA system based on the adaptive Gaussian approximation.  Most existing works~\cite{LXPD-2015,CGN-2017,DPXYD-2016,YBZJPZ-2016} focus on either optimizing the procedures of MD constellation construction or improving the performance of MPA under the framework in \cite{MHAH-2014}. However, the practical utilization of NOMA systems can be further improved if joint multi-user codeword design can be achieved.


In this paper, we jointly design codewords for multiple users simultaneously via MD trellis-coded modulation (TCM) techniques. The data streams of different users are jointly coded and mapped directly to the sparse codewords. Traditionally, TCM techniques~\cite{SD-2004} have been well investigated as a suitable approach to jointly optimize the error control coding and modulation through the signal set expansion instead of bandwidth expansion. Differently, we consider the MD-TCM technique purely as a method to design codewords for NOMA users, instead of replacing the channel coding. New challenges have been posed on both the codeword design and the multi-user detection. On one hand, due to the multiplexing nature, the mapping from the MD constellation point to the superimposed codeword is different from that in the traditional MD-TCM scheme. Therefore, a new method for selecting and labeling the MD constellation should be considered. On the other hand, due to the entanglement of encoding and multiplexing, the traditional MPA or Viterbi algorithm cannot be directly applied to decode the TCM-based NOMA signals, requiring a new decoding scheme.


To tackle the above challenges, we design new encoding and decoding schemes for joint codeword design in an overloaded NOMA system to improve the coding gain. Over each subcarrier, the data streams of multiple users are rearranged into one data sequence, which is then coded and mapped to an MD constellation point. Our main contributions in this paper can be summarized as below:
\begin{itemize}
\item To construct the TCM encoder, a novel bipartite set partitioning algorithm based on farthest-point optimization (FPO)~\cite{TDO-2011} is proposed and analyzed such that the minimum free squared Euclidean (MFSE) distance of the system can be maximized.
\item A maximum likelihood sequence detection (MLSD) is developed. To balance the complexity and the BER performance, we also propose a suboptimal two-layer joint decoding scheme based on the Viterbi algorithm. The non-orthogonal nature of NOMA is utilized in this scheme and the decoding complexity is analyzed.
\item Simulation results show that the proposed scheme significantly outperforms the traditional code-domain NOMA and OFDMA in terms of the BER performance. The influence of system parameters on the BER performance and decoding complexity is also investigated.
\end{itemize}

The rest of this paper is organized as follows. In Section \uppercase\expandafter{\romannumeral2}, we introduce the framework of TCM-based NOMA. In Section \uppercase\expandafter{\romannumeral3}, we discuss the TCM-based NOMA encoder design criteria and detailed steps. Based on the criteria, an FPO-based bipartite set partitioning algorithm is proposed and analyzed. In Section \uppercase\expandafter{\romannumeral4}, we provide an optimal MLSD decoding scheme. To balance the complexity and BER performance, we also develop a suboptimal two-layer soft-decision based Viterbi algorithm. Simulation results are presented in Section \uppercase\expandafter{\romannumeral5}, and finally, we conclude the paper in Section \uppercase\expandafter{\romannumeral6}.

\section{System Model}
In this section, we first introduce downlink code-domain NOMA and then present the key idea of TCM-based joint NOMA codeword design.
\subsection{Code-Domain NOMA Multiplexing}
\begin{figure*}[!t]
\centering
\includegraphics[width=5.5in]{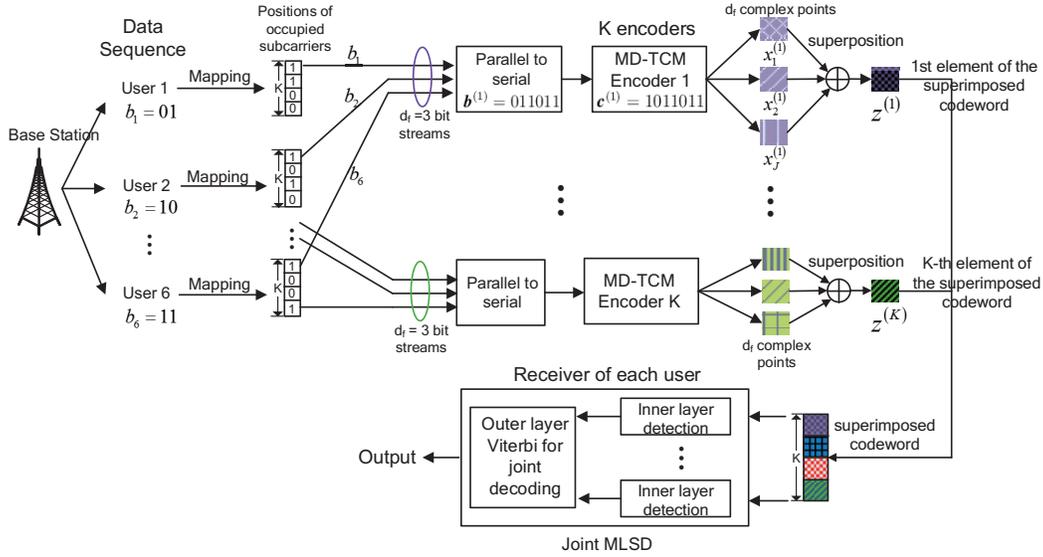}
\vspace{-0.2cm}
\caption{The general structure of a code-domain NOMA system with joint coded modulation in which $J = 6$, $K = 4$, $N = 2$, $d_f = 3$, and the mapping matrix is shown in equation (2).} \label{system_model}
\vspace{-0.3cm}
\end{figure*}

The general structure of a code-domain NOMA system is shown in Fig.~\ref{system_model}. In the figure, the BS sends $J$ data streams, each to a NOMA user. The available bandwidth is divided into $K$ orthogonal subcarriers. Different from orthogonal access, multiple users can share the same subcarrier simultaneously.

At the transmitter, the data stream\footnote{In Fig.~\ref{system_model}, we omit the subscript $t$ due to the limited space.} of each user $j$ in time unit $t$, denoted by ${{\textbf{\emph{b}}}_j}\left[ t \right]{\rm{ = }}\left( {{b_{j}^1}\left[ t \right],\cdots, {b_{j}^q}\left[ t \right]} \right)$, is transmitted over $N$ ($N \ll K$) subcarriers. Data stream ${\textbf{\emph{b}}_j}\left[ t\right]$ is mapped to a sparse codeword ${{\textbf{\emph{x}}}_j}\left[ t \right]{\rm{ = }}\left( {x_j^{\left( 1 \right)}\left[ t \right], \cdots ,x_j^{\left( K \right)}\left[ t \right]} \right)$ of length $K$, in which $N$ nonzero elements of this codeword represent the set of intended signal points and $(K -N)$ zero elements correspond to the unoccupied subcarriers of user $j$. All $J$ NOMA users are multiplexed over $K$ shared subcarriers. Therefore, the received signal over subcarrier $k$ can be expressed as
\begin{equation} \label{received1}
{y^{\left( k \right)}\left[ t\right]} = {h^{\left( k \right)}\left[ t\right]}\sum\limits_{j = 1}^J {x_j^{\left( k \right)}\left[ t\right] }  + {n^{\left( k \right)}\left[ t\right] },
\end{equation}
where the channel coefficient of subcarrier $k$ in time unit $t$ is denoted by $h^{\left( k \right)}\left[ t\right] $ and the additive white Gaussian noise~(AWGN) is denoted as ${n^{\left( k \right)}\left[ t\right] } \sim \mathcal{CN}\left( {0,{\sigma}^2} \right)$, with ${\sigma}^2$ as the noise variance.

\subsection{Mapping Matrix Design}
The occupied subcarrier set of each user is fixed, that is, for each user the positions of non-zero elements in the sparse codeword with respect to any data streams are the same. We can use a binary $K \times J$ matrix $\emph{\textbf{F}}$ to depict such mapping relation where each variable $f_{k,j}$ in $\emph{\textbf{F}}$ indicates whether user $j$ occupies subcarrier $k$, i.e., whether $x_j^{\left( k \right)}\left[ t\right]$ is a non-zero element.

Since each codeword contains $N$ non-zero elements out of $K$, there are $C_K^N$ possible mappings between the subcarriers and each user. Note that different users occupy different subset of subcarriers. Therefore, at most $J = C_K^N$ users can be supported simultaneously. Denote $d_f$ as the number of users sharing the same subcarrier in this case. We then have $d_f = C_{K - 1}^{N - 1}$.

For the NOMA system in Fig.~\ref{system_model} where $J = 6$, $K = 4$, $N = 2$, and $d_f = 3$, a widely used mapping matrix is presented as below:
\begin{equation} \label{mapping_matrix}
\emph{\textbf{F}} = \left[ {\begin{array}{*{20}{c}}
   1 & 1 & 0 & 0 & 0 & 1  \\
   1 & 0 & 0 & 1 & 1 & 0  \\
   0 & 1 & 1 & 1 & 0 & 0  \\
   0 & 0 & 1 & 0 & 1 & 1  \\
\end{array}} \right]
\end{equation}
In Fig.~\ref{system_model}, we use a square marked by 1 to represent an occupied subcarrier by each user, otherwise a blank square marked by 0. With the mapping matrix $\emph{\textbf{F}}$, we can then denote the set of users occupying subcarrier $k$ as ${{\cal{J}}^{\left( k \right)}} = \left\{ {j|1 \le j \le J,{f_{k,j}} = 1} \right\}$. For example, for the mapping matrix in $\left( \ref{mapping_matrix} \right)$, we have ${{\cal{J}}^{\left( 1 \right)}} = \left\{1, 2, 6\right\}$.

\subsection{TCM-based Joint NOMA Codeword Design}
In the traditional NOMA scheme~\cite{HH-2013}, the codeword of each user, $\emph{\textbf{x}}_j\left[ t\right] $, is only related to its data stream $\textbf{\emph{b}}_j\left[ t\right] $. Each user independently selects a codeword from a pre-defined codebook. Though such a design provides a straightforward decoding method, it greatly limits the potential coding gain due to the independent coding of each user. In addition, channel coding needs to be performed ahead of the MD modulation in the existing scheme. Therefore, extra bandwidth expansion is required when both the channel coding and modulation are considered, leading to inefficient utilization of frequency resources.

To tackle the above two issues, we jointly determine the non-zero elements of all the users over each subcarrier based on their data streams. We consider a joint coding and modulation scheme MD-TCM in which the binary convolutional code and M-ary constellation are combined to obtain the coding gain through the signal set expansion instead of the bandwidth expansion. The key idea can be illustrated as below.
\subsubsection{TCM-based NOMA Encoding}
Given the set of users occupying subcarrier $k$, ${{\cal{J}}^{\left( k \right)}}$, the data sequence transmitting over subcarrier $k$ consists of data streams from $d_f$ users in time unit $t$, which can be denoted as ${\emph{\textbf{b}}^{(k)}\left[ t\right] } = \left\{ {{\emph{\textbf{b}}_j\left[ t\right] }|j \in {\cal{J}}^{\left( k \right)}} \right\}$. As mentioned above, according to the first row of the mapping matrix $\textbf{\emph{F}}$ in equation $\left( \ref{mapping_matrix} \right)$, we have ${{\cal{J}}^{\left( 1 \right)}} = \left\{1, 2, 6\right\}$. Therefore, the data streams for user 1, 2, and 6 in the current time slot $t$ in Fig.~\ref{system_model}, i.e., ${\emph{\textbf{b}}_1}\left[ t \right] = 01$, ${\emph{\textbf{b}}_2}\left[ t \right] = 10$, and ${\emph{\textbf{b}}_6}\left[ t \right] = 11$, respectively, are transmitted over subcarrier 1. Thus, we have ${\emph{\textbf{b}}^{\left( 1 \right)}}\left[ t \right] = 011011$.


In each time unit $t$, a data sequence ${\emph{\textbf{b}}^{(k)}}\left[ t\right] $ is encoded into a sequence ${\emph{\textbf{c}}^{(k)}}\left[ t\right]$ and then mapped to an MD constellation point via the MD-TCM technique such that the event-error probability can be minimized. Each component of the MD constellation point represents a non-zero element of the sparse codewords over subcarrier $k$, i.e., $x_j^{\left( k \right)}$, $j \in {\cal{J}}^{\left( k \right)}$. In Fig.~\ref{system_model}, after passing the MD-TCM encoder, the data sequence ${{\bf{b}}^{\left( 1 \right)}}\left[ t \right] = \left( {011011} \right)$ is first encoded as ${{\bf{c}}^{(1)}}\left[ t \right] = \left( {1011011} \right)$ and then projected onto a 6D constellation point consisting of three 2D components, i.e., $x_1^{\left( 1 \right)}\left[ t \right]$, $x_2^{\left( 1 \right)}\left[ t \right]$, and $x_6^{\left( 1 \right)}\left[ t \right]$ in Fig.~\ref{system_model}.


The $k$-th element of the superimposed codeword in time unit $t$ can be obtained by adding these 2D components $\sum\nolimits_{j \in {{\cal{J}}^{\left( k \right)}}} x_j^{\left( k \right)}\left[ t\right] $, carrying the information of all users in ${{\cal{J}}^{\left( k \right)}}$. The superimposed codeword is then determined after all $K$ elements are obtained through the above process. In Fig.~\ref{system_model}, the first element of the superimposed codeword is $\left( {x_1^{\left( 1 \right)}\left[ t \right] + x_2^{\left( 1 \right)}\left[ t \right] + x_6^{\left( 1 \right)}\left[ t \right]} \right)$. Other elements in the superimposed codeword can be similarly obtained, represented by the dark squares in a vertical column.

\subsubsection{TCM-based NOMA Decoding}
At the receiver of each user, we jointly decode the data streams of all users by utilizing the soft-decision Viterbi based MLSD. Depending on the mapping matrix $\emph{\textbf{F}}$, the same data stream of each user transmits over different subcarriers, which can be utilized for joint decoding and is different from the traditional Viterbi algorithm.

In each step of the sequence detection, we first estimate the transmitted signal points by utilizing the above non-orthogonal nature. We then update the survivor paths based on the soft-decision Viterbi algorithm. The data streams of each user can be recovered based on the survivor paths.

%

\section{TCM-based NOMA Encoder Design}
In this section, we first discuss the criteria for TCM-based NOMA encoder design. and then illustrate three phases of the encoder design in detail.

\subsection{Encoder Design Criteria}
\subsubsection{Notations}
We assume that the number of subcarriers is $K = 2^p$, and the number of bits to be coded and transmitted for each user $j$ in each time unit $t$ is $q = |\emph{\textbf{b}}_j\left[ t\right] |$, $1 \leq j \leq J$, $\forall t > 0$. For example, in Fig.~\ref{system_model}, we have $q = 2$ for each user. The sequence of non-zero elements in the sparse codewords with respect to subcarrier $k$ is denoted as $\emph{\textbf{x}}^{\left( k \right) }\left[ t\right]  = \left\{ {x_j^{\left(k \right) }\left[ t\right] }|j \in {\cal{J}}^{\left( k \right)} \right\}$.

\begin{figure}[!t]
\centering
\includegraphics[width=3.5in]{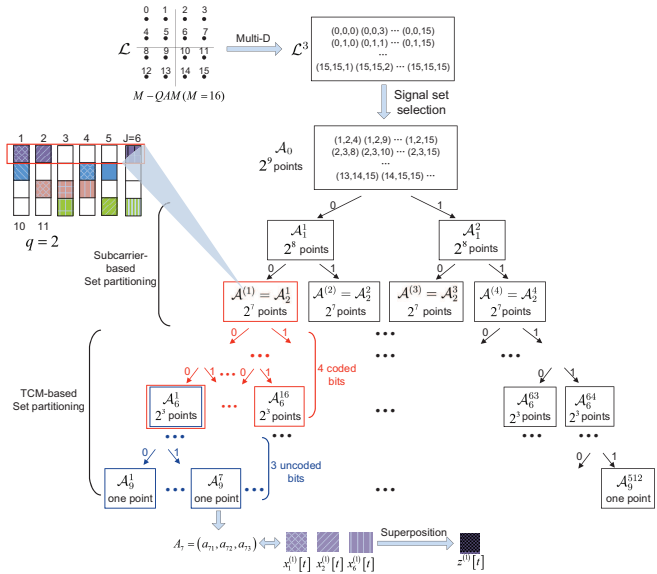}
\vspace{-0.7cm}
\caption{MD-TCM based NOMA encoder diagram with $J = 6$, $K = 4$, $d_f = 3$, $M = 16$, $r = 3$, $p = 2$, and $q = 2$.} \label{MD-TCM}
\vspace{-0.5cm}
\end{figure}

As shown in Fig.~\ref{MD-TCM}, we utilize the conventional lattice or star constellation to construct an MD mother constellation from which we select the signal set. To be specific, given a M-QAM constellation $\mathcal{L}$ with size $M$, a $d_f \times$M-QAM mother constellation ${\mathcal{L}^{{d_f}}}$ can be constructed with the size of $M^{d_f}$. In Fig.~\ref{MD-TCM}, we have a $3\times$16QAM mother constellation ${\mathcal{L}^{3}}$.

%
%
%
%

Since the original constellation $\mathcal{L}$ is composed of 2D signal points, the MD mother constellation consists of $2d_f$-D points. Each $2d_f$-D point $A_l \in {\mathcal{L}^{{d_f}}}$ is denoted by ${A_l} = \left( {{a_{l1}},{a_{l2}}, \cdots, {a_{ld_f}}} \right)$ with $1 \le l \le {M^{d_f}}$, in which $a_{li}$ represents a signal point from the original M-QAM constellation $\mathcal{L}$. We denote the position of a $2d_f$-D point $A_l$ in the 2D plane\footnote{For convenience, we denote the position $z_l$ of the signal point $A_l$ in the form of a complex number. The coordinates of this point in the 2D plane is $({\mathop{\rm Re}\nolimits} \left( {{z_l}} \right), {\mathop{\rm Im}\nolimits} \left( {{z_l}} \right))$. In the remaining part of this paper, when we mention the position of a $2d_f$-D point $A_l$, we refer to its position in the 2D plane.} as $z_l = \sum\nolimits_{i = 1}^{{d_f}} {{a_{li}}}$. For example, if we have three 2D points extracted from $\cal{L}$, say, $\left( {1 + 1i} \right)$, $\left( {-3 + 3i} \right)$, and $\left( {-1 - 1i} \right)$ where $i = \sqrt { - 1}$, and then a 6D point $A_1$ in Fig.~\ref{MD-TCM} constructed from these points is denoted by ${A_7} = \left( {1 + 1i,-3 + 3i,-1 - 3i} \right)$, and its position in the 2D plane is $z_7 = \left(-3 + 1i\right)$.

\subsubsection{Criteria}
With the above definitions, the process of coded modulation can be described as below. According to the TCM principle, in each time unit $t$, $r$ out of $q{d_f}$ bits from the data sequence ${\emph{\textbf{b}}^{(k)}}\left[ t\right] $ are sent into a convolutional encoder of rate $r/\left( r + 1 \right)$. The rest of $qd_f - r$ uncoded bits from ${\emph{\textbf{b}}^{(k)}}\left[ t\right] $ will determine a specific point in a subset of ${\mathcal{L}^{{d_f}}}$ with the size of ${2^{q{d_f} - r}}$. In this way, the coded sequence ${\emph{\textbf{c}}^{(k)}\left[ t\right]}$ with the length of $q{d_f} + 1$ is mapped to a unique $2d_f$-D constellation point $A_l \in {\mathcal{L}^{{d_f}}}$. The $k$th element in the superimposed codeword can be determined based on the position of a chosen $2d_f$-D point $A_l$, i.e., $z^{\left( k\right) } = {\sum _{j \in {{\cal{J}}^{\left( k \right)}}}}{x_j}^{\left( k \right) }\left[ t\right] = \sum\nolimits_{i = 1}^{{d_f}} {{a_{li}}}$.

Take Fig.~\ref{MD-TCM} as an example. For subcarrier 1, the data sequence to be transmitted is ${\emph{\textbf{b}}^{\left( 1 \right)}}\left[ t \right] = 011011$. Given a signal constellation ${\cal{A}}^{\left( 1\right)}$ of size 128, the first three bits of this sequence are sent into a convolutional encoder of rate 3/4 such that the output four coded bits, say, 1011, select a signal subset of size 8, say, ${\cal{A}}_6^2$. A specific 6D point of ${\cal{A}}_6^2$, say, $A_7$, is then selected based on the remaining three uncoded bits of ${\emph{\textbf{b}}^{\left( 1 \right)}}\left[ t \right]$. The value of the first element in the superimposed codeword is actually the position of $A_7$. The mapping between the coded (or uncoded) bits and the signal subset (or a 6D point) will be illustrated in Section III.B-2.

To minimize the event-error probability~\cite{SD-2004}, we aim to jointly design the coding and modulation so that the MFSE distance between any two coded sequences, denoted by $d_{free}^2$, can be maximized. In other words, we aim to construct an optimal mapping from the transmitted data sequence over each subcarrier to the corresponding non-zero elements in the codewords, i.e., ${\emph{\textbf{b}}^{\left( k \right)}} \to {\emph{\textbf{c}}^{(k)}} \to {A_l} \to \emph{\textbf{x}}^{\left( k \right) }$, $1 \le k \le K$, $A_l \in {\mathcal{L}^{{d_f}}}$. Since the mapping is not related to $t$, we omit it in the remaining part of this section.

As mentioned above, not all bits of sequence ${\emph{\textbf{b}}^{(k)}}$ participate in the convolution computation, and thus, there exist parallel branches between two states of the convolutional encoder, brought by the uncoded bits. According to the TCM principle, $d_{free}^2$ is determined by two terms: 1) the minimum squared Euclidean (MSE) distance $\delta _{free}^2$ between different trellis paths longer than one branch; 2) the MSE distance $\delta _{\min }^2$ between the parallel branches of the encoder. Therefore, we have $d_{free}^2 = \min \left\{ {\delta _{free}^2,\delta _{\min }^2} \right\}$. In Subsections III.B and III.C, we illustrate how to calculate $\delta _{\min }^2$ and $\delta _{free}^2$, respectively, as well as the relationship between them.

\subsection{Joint Encoding}
Three phases are illustrated as below, i.e., 1) signal set selection; 2) signal set labeling; 3) convolutional encoder construction.
\subsubsection{Signal Set Selection}
A signal set is selected from the MD mother constellation so as to satisfy the following two constraints on a) the size of the set; b) the uniqueness of each point's position in the set. For successful decoding, each coded data sequence $\emph{\textbf{c}}^{\left( k\right) }$ of length $(q{d_f} + 1)$ is required to be mapped to a unique position. For generality, we also assume that any two coded data sequences $\emph{\textbf{c}}^{\left( k\right) }$ and $\emph{\textbf{c}}^{\left( k'\right) }$ transmitted over different subcarriers $k$ and $k'$ are mapped to two different positions\footnote{In the simulation, we will show both cases where different signal sets or identical signal sets are allocated to different subcarriers.}. Since there are $K$ subcarriers, ${2^{p + q{d_f} + 1}}$ unique positions in total are required for the selected signal set.

Therefore, we aim to construct a signal set ${\cal{A}}_0$ of size ${2^{p + q{d_f} + 1}}$ from the MD mother constellation ${\mathcal{L}^{{d_f}}}$ such that the positions of any two points projecting in the 2D plane are different, which can be mathematically formulated as:
\begin{subequations} \label{signal_set_selection}
\begin{align}
\mbox{find}\ & {{{\cal{A}}_0}} \label{system_1}\\
\textbf{\emph{s.t.}}\ & {z_l} \ne {z_{l'}}, \forall {A_l},{A_{l'}} \in {{\cal{A}}_0}, l \ne l',\label{system_2}\\
& \left| {{{\cal{A}}_0}} \right| = {2^{p + q{d_f} + 1}}, \label{system_3}\\
& {\cal{A}}_0 \subseteq {\mathcal{L}^{{d_f}}}. \label{system_4}
\end{align}
\end{subequations}
This problem can be solved by the following two steps.

\textbf{a) Obtaining a unique constellation:}
Note that in the MD mother constellation ${\mathcal{L}^{{d_f}}}$, there may exist more than one $2d_f$-D constellation points sharing the same position in the 2D plane. Denote the set of points sharing the same positions as ${\cal{A}}_s$. To keep the uniqueness of each signal point, we can only select one point $A_{l^*} \in {\cal{A}}_s$ as a member of ${\cal{A}}_0$ while all the other points are removed. To improve the diversity among $d_f$ non-zero elements of the sparse codewords transmitting over the same subcarrier, we select the MD point with the largest variance among its 2D components, i.e., ${l^*} = \arg \mathop {\max }\limits_{{A_l} \in {{\cal{A}}_s}} {\mathop{\rm var}} \left( {{A_l}} \right)$. Note that the size of ${\cal{A}}_0$ is required to be no smaller than ${2^{p + q{d_f} + 1}}$; otherwise, the MD mother constellation ${\mathcal{L}^{{d_f}}}$ is required to be reconstructed.

\textbf{b) Shaping:}
After step a), we denote the current constellation as ${\cal{A}}_c$. If the size of ${\cal{A}}_c$ is larger than ${2^{p + q{d_f} + 1}}$, we continue to remove points until constraint $\left( \ref{system_3} \right)$ is satisfied. For a constellation, its average energy is required to be minimized while the MFSE distance between points can be as large as possible. Therefore, we tend to remove those points close to their neighbors but far away from the original point. A greedy algorithm is adopted for shaping based on the above criterion. We sort the constellation points in ${\cal{A}}_c$ as ${A_{\pi \left( 1 \right)}},{A_{\pi \left( 2 \right)}}, \cdots$ such that
\begin{equation} \label{shaping_relation}
\frac{{\mathop {\min }\limits_{{A_r} \in {{\cal{A}}_c},r \ne \pi \left( j \right)} \left[ {{z_{\pi \left( j \right)}} - {z_r}} \right]}}{{{{\left| {{z_{\pi \left( j \right)}}} \right|}^2}}} \le \frac{{\mathop {\min }\limits_{{A_r} \in {{\cal{A}}_c},r \ne \pi \left( {j'} \right)} \left[ {{z_{\pi \left( {j'} \right)}} - {z_r}} \right]}}{{{{\left| {{z_{\pi \left( {j'} \right)}}} \right|}^2}}},
\end{equation}
holds $\forall j \le j'$, where the numerator in the fraction is the MFSE distance between point ${A_{\pi \left( j \right)}}$ and other points in ${\cal{A}}_c$, and the denominator represents the energy of point ${A_{\pi \left( j \right)}}$. We remove the points one by one in ${\cal{A}}_c$ starting from ${A_{\pi \left( 1 \right)}}$ until the size of the remained signal set is ${2^{p + q{d_f} + 1}}$, i.e., the target set ${\cal{A}}_0$ is obtained.



\subsubsection{Signal Set Labeling}



Signal set labeling refers to the process in which each coded sequence $\emph{\textbf{c}}^{\left( k\right) }$, $1 \le k \le K$ is assigned a unique signal point in ${\cal{A}}_0$ such that the MFSE distance can be maximized. As shown in Fig.~\ref{MD-TCM}, two steps of signal set labeling for NOMA are listed as below:
\begin{itemize}
\item \textbf{Subcarrier-based Set Partitioning:} As mentioned above, any two coded sequences $\emph{\textbf{c}}^{\left( k\right)}$ and $\emph{\textbf{c}}^{\left( k' \right)}$ are assigned to different signal points in ${\cal{A}}_0$. Therefore, we divide ${\cal{A}}_0$ into $K$ subsets with the equal size, i.e., ${\mathcal{A}_0} \to \left\{ {{{\cal A}^{\left( 1 \right)}},{{\cal A}^{\left( 2 \right)}}, \cdots ,{{\cal A}^{\left( K \right)}}} \right\}$. Each subset ${\mathcal{A}^{\left( k\right) }}$ corresponds to all possible signal points transmitting over subcarrier $k$, satisfying $\left| {{{\cal{A}}^{\left( k\right) }}} \right| = {2^{q{d_f} + 1}}$.
\item \textbf{TCM-based Set Partitioning:} For each subcarrier $k$, a signal point in ${\cal{A}}_k$ is assigned to a coded sequence $\emph{\textbf{c}}^{\left( k\right) }$ such that the MFSE distance can be maximized. To construct the mapping based on the above criterion, we adopt the set partitioning technique in which a binary partitioning tree is utilized~\cite{SD-2004}. In each level of the tree, the current signal set is divided into two subsets such that the minimum squared subset distance (MSSD), that is, the MSE distance between signal points in the same subset, is maximized. In the last level of the tree, the leaf node refers to a constellation point. Each constellation point can be reached via a unique path through the tree.
\end{itemize}

\textbf{Overall Set Partitioning Process:}
To further maximize the MFSE distance of the coding scheme with respect to each subcarrier, we consider adopting the binary-tree based set partitioning technique in the first step as well. This is to say, given ${\cal{A}}_0$, we aim to maximize the MSSD of each subset ${\mathcal{A}^{\left( k\right) }}$. For the system shown in Fig.~\ref{system_model} and Fig.~\ref{MD-TCM}, the signal set labeling can be described as below. We first construct a super binary partitioning tree of ${2^9}$ levels with the root node ${\cal{A}}_0$. The first level of the tree consists of two subsets divided from ${\cal{A}}_0$, denoted by ${\cal{A}}_1^1$ and ${\cal{A}}_1^2$. For level $2 \le j \le {2^{9}}$ of the tree, each branch node $A_j^i$ represents a point set which will be divided into two subsets in the next level, i.e., ${\cal{A}}_{j + 1}^{2i - 1}$ and ${\cal{A}}_{j + 1}^{2i}$. The signal sets obtained after the subcarrier-based set partitioning can be found in level 2 of the tree, i.e., $\{{\mathcal{A}^{\left( 1\right) }}, \cdots, {\mathcal{A}^{\left( 4\right) }}\} = \left\{ {{\cal{A}}_2^1, \cdots ,{\cal{A}}_2^{{4}}} \right\}$. In the last level of the tree, we have ${\cal{A}}_{9}^l = \left\{ {{A_l}} \right\}, {A_l} \in {\cal{A}}_0$, i.e., each leaf node represents a specific signal point. The coded sequence mapped to this point is labeled by the path towards this leaf node. For example, we have ${\cal{A}}_{9}^1 = \left\{ {{A_1}} \right\}$ implying that the coded sequence 0000000 transmitted over subcarrier 1 is mapped to the signal point $A_1 \in {\cal{A}}^{\left(1\right)}$.

Based on the binary subtree for each subcarrier $k$, ${\mathcal{A}^{\left( k\right) }}$, we observe that $\delta _{\min }^2$ can be obtained by inspecting the minimum MSSD of the subsets in the $(r+1)$th level of the subtree, i.e., the $(p+r+1)$th level of the binary tree in Fig.~\ref{MD-TCM}.
\begin{equation} \label{delta_min}
\delta _{\min }^2 = \mathop {\min }\limits_{1 \le i \le {2^{r + 1}}} \left\{ {\Delta \left( {{\cal{A}}_{p + r + 1}^{\left( {k - 1} \right) \cdot {2^{r + 1}} + i}} \right)} \right\}.
\end{equation}
%

\textbf{Basic Bipartite Set Partitioning Operation:} Note that the basic operation in the above process is to divide a signal set into two subsets with the equal size. We now formulate this basic bipartite set partitioning problem and propose a novel algorithm to solve it.

\textbf{a) Problem Formulation:}
Given a signal set ${\cal{A}}_b$, we aim to divide it into two subsets ${\cal{A}}_b^1$ and ${\cal{A}}_b^2$ with equal size such that the MSSD is maximized. The MSSD of each subset ${\cal{A}}_b^j$ is given by
\begin{equation} \label{MSSD1}
\Delta \left( {{\cal{A}}_b^j} \right) = \mathop {\min }\limits_{l,l' \in {\cal{A}}_b^j,l \ne l'} \left\| {z_l  - z_{l'} } \right\|_2^2,j = 1,2.
\end{equation}
Therefore, an optimal bipartite set partitioning of ${\cal{A}}_b$, i.e., ${\cal{A}}_b \to \{ {\cal{A}}_b^1, {\cal{A}}_b^2 \}$, can be obtained by solving the following problem:
\begin{subequations} \label{bipartite_partition}
\begin{align}
& \mathop {\max }\limits_{\left\{ {{\cal{A}}_b^1,{\cal{A}}_b^2} \right\}} \left[ {\mathop {\min }\limits_{1 \le j \le 2} \Delta \left( {{\cal{A}}_b^j} \right)} \right]  \label{partition_1}\\
\textbf{\emph{s.t.}}\ & \left| {{\cal{A}}_b^j} \right| = \frac{1}{2}\left| {{{\cal{A}}_b}} \right|, j = 1, 2,\label{partition_2}\\
& {\cal{A}}_b^j \subset {{\cal{A}}_b}, j = 1, 2, \label{partition_3}\\
& \Delta \left( {{\cal{A}}_b^j} \right) > 0, j = 1, 2. \label{partition_4}
\end{align}
\end{subequations}
This is a non-trivial problem due to its combinatorial nature and the irregular positions of the points. The traditional multi-level coding technique~\cite{SRAGD-1990} which is originally designed for partitioning lattice or star constellations does not fit this case any more. We then propose a modified farthest point optimization algorithm to address this problem, as illustrated below.

\textbf{b) Algorithm Design:}
We observe that the distribution of the points in our formulated problem has the \emph{blue noise} properties, i.e., i) the signal points cover a certain area and there are no ``holes" or ``clusters" in the 2D plane; ii) the points in the selected set are distributed almost irregularly. Such properties have drawn great attention in the researches on point set generation~\cite{ZZYLW-2012} where a farthest point optimization (FPO) strategy~\cite{TDO-2011} has been utilized for generating point distributions with high-quality ``blue noise" characteristics, i.e., large point spacing in a given area.

Instead of point generation, we aim to select points for each subset given the point distribution. By extending the FPO strategy, we then propose a novel bipartite set partitioning (FPO-BSP) algorithm to solve problem $\left(\ref{bipartite_partition}\right)$, consisting of two phases: i) initial subset construction; ii) FPO iteration. Details of the algorithm are shown in Algorithm \ref{Alg_1} and are illustrated as below.

For convenience, we first present three different distance metrics. Given a signal point $A_l$, its minimum distance to a set ${\cal{A}}_x$ is defined as
\begin{equation}
{d_l}\left( {{{\cal{A}}_x}} \right) = \mathop {\min }\limits_{{A_r} \in {{\cal{A}}_x},r \ne l} \left| {{z_l} - {z_r}} \right|.
\end{equation}
Based on this definition, the MSSD of a set ${\cal{A}}_x$ in $\left( \ref{MSSD1} \right)$ can be rewritten by
\begin{equation} \label{MSSD2}
\Delta \left( {{{\cal{A}}_x}} \right) = \mathop {\min }\limits_{{A_l} \in {{\cal{A}}_x}} {d_l}\left( {{{\cal{A}}_x}} \right).
\end{equation}
The average minimum distance of a set ${\cal{A}}_x$, denoted by $\bar d\left( {{A_x}} \right)$, can be obtained based on each point's minimum distance, i.e.,
\begin{equation} \label{average_minimum_distance}
\bar d\left( {{{\cal{A}}_x}} \right) = \frac{1}{{\left| {{{\cal{A}}_x}} \right|}}\sum\limits_{{A_l} \in {{\cal{A}}_x}} {{d_l}\left( {{{\cal{A}}_x}} \right)}.
\end{equation}

Since we tend to divide the target signal set into two subsets where the points are spread out as far as possible, both the MSSD and the average minimum distance of the subsets are encouraged to be maximized.


\textbf{i) Initial Subset Construction:} As shown in Phase 1 of Algorithm 1, we adopt a greedy method to divide the target set into two initial subsets. For the target set ${\cal{A}}_b$ of size $\left| {\cal{A}}_b \right|$, there are $L = C_{\left| {\cal{A}}_b \right|}^2$ pairs of points in total. We sort these pairs in the increasing order of the distance between two points (line 2 in Algorithm \ref{Alg_1}). The list $\Gamma$ can be mathematically presented as
\begin{equation} \label{Sorting_list}
{\Gamma} = \left\{ {\left( {{A_{{\alpha} (1)}},{A_{\beta \left( 1 \right)}}} \right),\left( {{A_{\alpha \left( 2 \right)}},{A_{\beta \left( 2 \right)}}} \right), \cdots ,\left( {{A_{\alpha \left( L \right)}},{A_{\beta \left( L \right)}}} \right)} \right\},
\end{equation}
satisfying that
\begin{equation} \label{List_distance}
\left| {{z_{\alpha \left( j \right)}} - {z_{\beta \left( j \right)}}} \right| \le \left| {{z_{\alpha \left( {j'} \right)}} - {z_{\beta \left( {j'} \right)}}} \right|,1 \le j \le j' \le L.
\end{equation}

As shown in line 3 of Algorithm \ref{Alg_1}, we initialize two subsets as ${\cal{A}}_b^1 = \left\{ {{A_{\alpha \left( 1 \right)}}} \right\}$ and ${\cal{A}}_b^2 = \left\{ {{A_{\beta \left( 1 \right)}}} \right\}$. The remaining pairs of points in list $\Gamma$ are then traversed to be added into different subsets (line 5-21). To be specific, we consider a pair ${\left( {{A_{\alpha \left( j \right)}},{A_{\beta \left( j \right)}}} \right)}$. If both points never show up in the current subsets, we then add ${{A_{\alpha \left( j \right)}}}$ and ${{A_{\beta \left( j \right)}}}$ separately into ${\cal{A}}_b^{\tilde \alpha }$ and ${\cal{A}}_b^{\tilde \beta }$ in which $\tilde \alpha, \tilde \beta \in {{\mathbb{Z}}^ + }$ such that the minimum distance between the members of this pair and the subsets can be maximized (line 15-17), i.e.,
\begin{equation} \label{Sorting_inequality}
\begin{split}
&\min \left\{ {{d_{\alpha \left( j \right)}}\left( {{\cal{A}}_b^{\tilde \alpha }} \right),{d_{\beta \left( j \right)}}\left( {{\cal{A}}_b^{\tilde \beta }} \right)} \right\} \ge \min \left\{ {d_{\alpha \left( j \right)}}\left( {{\cal{A}}_b^{\tilde \beta }} \right), \right. \\
&\left. {d_{\beta \left( j \right)}}\left( {{\cal{A}}_b^{\tilde \alpha }} \right) \right\}, \tilde \alpha  + \tilde \beta  = 3.
\end{split}
\end{equation}
However, when the distance between ${{A_{\alpha \left( j \right)}}}$ and ${{A_{\beta \left( j \right)}}}$ is larger than the MSSD of current subsets, i.e., $\left| {{z_{\alpha \left( j \right)}} - {z_{\beta \left( j \right)}}} \right| > \max \left\{ {\Delta \left( {{\cal{A}}_b^1} \right),\Delta \left( {{\cal{A}}_b^2} \right)} \right\}$, then it is not necessary that ${{A_{\alpha \left( j \right)}}}$ and ${{A_{\beta \left( j \right)}}}$ are divided into two subsets. Each of them selects a subset with a large minimum distance to join (line 11-13). If one of the points in the target pair has already been added in a subset in previous operations, then the other point is naturally added to another subset (line 19-21). The whole process ends when the size of one subset reaches $\left| {\cal{A}}_b \right|/2$. The left unchosen points in list $\Gamma$ are then added to the other set which has not been fully filled (line 6-9).

\textbf{ii) FPO Iteration:}
Based on the above initialized subsets, we then perform the FPO iteration as shown in Phase 2 of Algorithm \ref{Alg_1}. The key idea of the FPO method is illustrated as below. Given two initial subsets ${\cal{A}}_b^1$ and ${\cal{A}}_b^2$, we consider replacing each point in subset ${\cal{A}}_b^1$ with the farthest position for this subset, which is selected from the other subset ${\cal{A}}_b^2$. The farthest position\footnote{Different from the researches on point distribution generating~\cite{TDO-2011}, we cannot generate new points in the given set, and thus, we redefine the farthest position as the farthest position in the existing points of the set. } for ${\cal{A}}_b^1$, defined as $A_{r^*} \in {\cal{A}}_b^2$, can be obtained by searching ${\cal{A}}_b^2$,
\begin{equation} \label{farther_position}
{r^*} = \arg \mathop {\max }\limits_{r \in {\cal{A}}_b^2} {d_r}\left( {{\cal{A}}_b^1} \right).
\end{equation}
For each point $A_l \in {\cal{A}}_b^1$, it is first removed from ${\cal{A}}_b^1$ and inserted into ${\cal{A}}_b^2$ (line 28). We then search for the farthest position for ${\cal{A}}_b^1\backslash \left\{ {{A_l}} \right\}$ from the set ${\cal{A}}_b^2 \cup \left\{ {{A_l}} \right\}$ based on equation $\left( \ref{farther_position} \right)$ (line 29-34). When all points in ${\cal{A}}_b^1$ are traversed once, one FPO iteration is finished (line 25-35). During multiple FPO iterations, both the MSSD and the average minimum distance of the subsets will be increasing until convergence, which will be proved in detail in Proposition 1.

\textbf{\emph{Delaunay Triangulation for FPO:}} Note that in the FPO method, a large amount of operations such as point searching, removing, and inserting are required. To achieve a low computational complexity, we introduce the Delaunay triangulation (DT) method to construct a dynamic graph in which the relative positions of points can be better depicted and easily traced.

\textbf{Definition 1:} Given a point set ${\cal{A}}_x$, a DT refers to a triangulation $D$ such that no point in ${\cal{A}}_x$ lies inside the circumcircle of any triangle in $D$. Any edge in a DT is called a Delaunay edge.

In a DT, each triangle follows the property of empty circumcircle. One commonly used method for constructing such a triangulation is the on-line DT method~\cite{LE-1992}. Starting from a certain point in the set, the neighboring points are inserted one by one to form the triangles while the property of empty circumcircle is guaranteed. Given a formulated triangulation $D$, point inserting and removing can be completed flexibly via the local optimization procedure in which Delaunay edges are added or removed within a localized area.

\textbf{Remark 1:} In Algorithm 1, we refer to the above operations as DT-INSERT $\left( D, A_l \right)$ and DT-REMOVE $\left( D, A_l \right)$, implying the point $A_l$ inserting into $D$ and removing from $D$, respectively.

\textbf{Remark 2:} In each FPO iteration, suppose that the DT constructed from the set ${\cal{A}}_x$ is $D$. The minimum distance between a point $A_l$ and ${\cal{A}}_x$ can be obtained by searching the Delaunay neighbors\footnote{The Delaunay neighbors of $A_l$ refer to those points in $D$ sharing the same edges with $A_l$.} of $A_l$ in $D$ instead of searching all points in ${\cal{A}}_x$. We refer to this operation as DT-SEARCH $\left( D, A_{l} \right)$.

\textbf{\emph{Algorithm Interpretation of Phase 2:}} Algorithm \ref{Alg_1} with respect to DT can be re-interpreted in detail as below. Following the on-line DT method, the triangulations of two initial point subsets ${\cal{A}}_b^1$ and ${\cal{A}}_b^2$ are constructed, denoted by $D_1$ and $D_2$. As shown in Phase 1, points are added into two subsets in sequence (line 4, 14, 18, 22). In each FPO iteration, for each vertex $A_l \in D_1$, the minimum distance between $A_l$ and ${\cal{A}}_b^1$, i.e., ${d_l}\left( {{{\cal{A}}_b^1}} \right)$, can be obtained by searching the Delaunay neighbors of $A_l$ in $D_1$ (line 26). After recording the current minimum distance ${d_{\max }} = {d_l}\left( {A_b^1} \right)$, vertex $A_l$ is removed from $D_1$ and inserted into $D_2$ (line 27-29). We then traverse the vertices in the newly constructed $D_2$ to search for the farthest position of $D_1$ (line 30-36). If the MSSD of ${\cal{A}}_b^2$ will not decrease and ${d_{{r^*}}}\left( {{\cal{A}}_b^1\backslash \left\{ {{A_l}} \right\}} \right)$ will not be smaller than the MSSD of ${\cal{A}}_b^1$, $A_l$ is then replaced by the farthest position of $D_1$ (line 35-37). The sizes of two subsets maintain to be equal since we just swap a new point in $D_2$ with an old one in $D_1$. The FPO iterations will not stop until no changes can be made to $D_1$ and $D_2$ (line 38).

\textbf{c) Analysis of the Proposed Algorithm:} The convergence and complexity of the FPO-BSP algorithm are analyzed as below. The proof of Proposition 1 can be found in Appendix A.

\textbf{Proposition 1:} In each FPO iteration of Algorithm \ref{Alg_1}, the average minimum distance and MSSD of subset ${\cal{A}}_b^1$ are increasing, and the MSSD of subset ${\cal{A}}_b^2$ are non-decreasing. Therefore, Phase 2 is guaranteed to converge.

\begin{algorithm}[!t]
	\caption{FPO-based bipartite set partitioning (FPO-BSP) algorithm}\label{Alg_1}
	
	\hspace*{-0.02in} {\bf Input:} 
	a point set ${\cal{A}}_b$ of size $N$\\
	\hspace*{-0.02in} {\bf Output:} 
	two subsets ${\cal{A}}_b^1$ and ${\cal{A}}_b^2$ of size $N/2$ with maximized MSSD
	\begin{algorithmic}[1]
		\State \textbf{Phase 1: Initial Subset Construction}
		\State Construct a list $\cal{L}$ based on equations $\left(\ref{Sorting_list}\right)$ and $\left(\ref{List_distance}\right)$.
		\State Set ${\cal{A}}_b^1 = \left\{ {{A_{\alpha \left( 1 \right)}}} \right\}$ and ${\cal{A}}_b^2 = \left\{ {{A_{\beta \left( 1 \right)}}} \right\}$.
		\State Initialize ${D_1} = {\left\{ {{A_{\alpha \left( 1 \right)}}} \right\}}$ and ${D_2} = {\left\{ {{A_{\beta \left( 1 \right)}}} \right\}}$.
		\For{$j = 2$ to $L$} 
		\If{$\left| {{\cal{A}}_b^1} \right| = N/2$ or $\left| {{\cal{A}}_b^2} \right| = N/2$}
		\State Add remained points to the subset not fully filled.
		\State break;
		\ElsIf{$\left\{ {{A_{\alpha \left( j \right)}},{A_{\beta \left( j \right)}}} \right\} \not\subset A_b^1 \cup A_b^2$}
		\If{$\left| {{p_{\alpha \left( j \right)}} - {p_{\beta \left( j \right)}}} \right| > \max \left\{ {\Delta \left( {{\cal{A}}_b^1} \right),\Delta \left( {{\cal{A}}_b^2} \right)} \right\}$}
		\State For each point $A_l \in \left\{ {{A_{\alpha \left( j \right)}},{A_{\beta \left( j \right)}}} \right\}$, set ${i^*} = \arg \mathop {\max }\limits_{i = 1,2} {d_l}\left( {{\cal{A}}_b^i} \right)$.
		\State $A_l$ joins the subset ${\cal{A}}_b^{i^*}$.
		\ElsIf{${{A_{\alpha \left( j \right)}}}, {{A_{\beta \left( j \right)}}} \notin {\cal{A}}_b^1 \cup {\cal{A}}_b^2$} 
		\State Find ${\tilde \alpha }$ and ${\tilde \beta }$ that satisfies inequality $\left( \ref{Sorting_inequality} \right)$.
		\State Add ${{A_{\alpha \left( j \right)}}}$ and ${{A_{\beta \left( j \right)}}}$ into ${\cal{A}}_b^{\tilde \alpha}$ and ${\cal{A}}_b^{\tilde \beta}$, respectively.
		\Else
		\State Set $A_l$ as the point in $\left\{ {{A_{\alpha \left( j \right)}},{A_{\beta \left( j \right)}}} \right\}$ which is not a member of ${\cal{A}}_b^1 \cup {\cal{A}}_b^2$.
		\State Add ${A_l}$ to the subset of which the other point is not a member, say, ${\cal{A}}_b^2$.
		\EndIf
		\EndIf
		\EndFor
		\State Construct two DTs $D_1$ and $D_2$.
		\State \textbf{Phase 2: FPO Iteration}
		\Repeat
		\For{each vertex $A_l$ in $D_1$}
		\State Obtain ${d_l}\left( {{\cal{A}}_b^1} \right)$ by DT-SEARCH $\left( D_1, A_l \right)$ in Remark 2
		\State Set \hspace{-0.1cm} $\left( {{A_{{r^*}}},{d_{\max }}} \right) = \left( {{A_l},{d_l}\left( {{\cal{A}}_b^1} \right)} \right)$
		\State Set ${\Delta _{\min }} = \Delta \left( {{\cal{A}}_b^2} \right)$
		\State DT-REMOVE$\left( {D_1}, {A_l} \right)$ and DT-INSERT$\left( {D_2}, {A_l} \right)$ based on Remark 1
		\For{each vertex $A_{l'}$ in $D_2$}
		\State DT-INSERT$\left( {D_1}, {A_{l'}} \right)$
		\State Obtain ${{d_{l'}}\left( {{\cal{A}}_b^1} \right)}$ by DT-SEARCH$\left( {D_1}, {A_{l'}} \right)$
		\State Obtain ${{d_{l}}\left( {{\cal{A}}_b^2} \right)}$ by DT-SEARCH$\left( {D_2}, {A_{l}} \right)$
		\State DT-REMOVE$\left( {D_1}, {A_{l'}} \right)$ 
		\If{${{d_{l'}}\left( {{\cal{A}}_b^1} \right)} > {d_{\max}}$ and ${d_l}\left( {{\cal{A}}_b^2} \right) \ge {\Delta _{\min }}$}
		\State Set $\left( {{A_{{r^*}}},{d_{\max }}} \right) = \left( {{A_{l'}},{d_{l'}}\left( {{\cal{A}}_b^1} \right)} \right)$
		\EndIf
		\EndFor
		\State DT-REMOVE$\left({A_{{r^*}}}, D_2\right)$,  DT-INSERT$\left({A_{{r^*}}}, D_1\right)$.
		\EndFor
		\Until {$D_1$ and $D_2$ do not change}
		\State \Return ${\cal{A}}_b^1$ and ${\cal{A}}_b^2$
	\end{algorithmic}
\end{algorithm}

\textbf{Remark 3:} Since the MSSD of two initial subsets of Phase 1 are usually different, we set the subset with a smaller MSSD as ${\cal{A}}_b^1$ and the other one as ${\cal{A}}_b^2$, and then send them to Phase 2. Based on Proposition 1, the MSSD of ${\cal{A}}_b^1$ and ${\cal{A}}_b^2$ will be more balanced after the iterations.

As discussed in Appendix B, the complexity of Phase 1 is $O\left( {{\left| {{{\cal{A}}_b}} \right|^2}\log \left| {{{\cal{A}}_b}} \right|} \right)$ and the complexity of each FPO iteration in Phase 2 is $O\left( {{\left| {{{\cal{A}}_b}} \right|}\log \left| {{{\cal{A}}_b}} \right|} \right)$.

\subsubsection{Convolutional Encoder Design}
As mentioned in Subsection II.A.2, a convolutional encoder is adopted in the TCM-based NOMA scheme to generate $\left( r + 1 \right)$ coded bits. We now illustrate how to design the rate $r/(r+1)$ and the structure of the convolutional encoder.

As shown in equation $\left( \ref{delta_min} \right)$ and Fig.~\ref{MD-TCM}, rate $r/(r+1)$ determines the level of set partitioning $p+r+1$ in a way that the value of ${\delta _{\min }^2}$ increases with $r$. Since the decoding complexity also grows with $r$, a trade-off should be reached between ${\delta _{\min }^2}$ and the complexity.
%

For the different rate $r/(r+1)$ with $1 \le r \le q{d_f} - 2$, the optimal convolutional encoder that maximizes $\delta _{free}^2$ is different. For simplicity, we assume that the same convolutional encoder of rate $r/(r+1)$ is designed for different subcarriers. The diagram of a systematic feedback convolutional encoder (Fig.~18-16 in \cite{SD-2004}) is adopted. Note that in the traditional MD-TCM scheme, the distance between any two nodes $A_l$ and $A_{l'}$ is $\sum\limits_{i = 1}^{{d_f}} {\left| {{z_{li}} - {z_{l'i}}} \right|}$. This can be considered as an upper bound of the distance between these two nodes in our proposed TCM-based NOMA scheme since we have
\begin{equation} \label{MD_TCM}
\sum\limits_{i = 1}^{{d_f}} {\left| {{z_{li}} - {z_{l'i}}} \right|}  \ge \left| {\sum\limits_{i = 1}^{{d_f}} {\left( {{z_{li}} - {z_{l'i}}} \right)} } \right|.
\end{equation}
Therefore, we adopt the structure of the convolutional encoder for ${d_f} \times$ M-PSK/QAM in the MD-TCM scheme~\cite{SD-2004}. For a given rate $r/(r+1)$, the value of $\delta _{free}^2$ increases with the number of register states, $V$, in the encoder. However, a large value of $V$ increases the decoding complexity as well, requiring a trade-off between $\delta _{free}^2$ and $V$. Since the optimal convolutional encoders have always been found by computer search~\cite{SRAGD-1990}, we do not illustrate the detailed process in this paper.

Note that the above method is suitable for the AWGN channels. However, for the Rayleigh fading, the number of parallel branches should be reduced and the encoder state diagram should be redesigned such that a) the shorted error event path length can be increased~\cite{DM-1988}; b) the product of the squared branch distance with respect to that path can be maximized~\cite{ST-1991}. Given an encoder designed for AWGN channels, we can modify the state diagram by reducing the parallel branches based on the above criteria.
\section{TCM-based NOMA Decoder Design}
\vspace{-0.1cm}
Due to the non-orthogonal nature of the NOMA, signals of multiple users are superimposed and transmitted over the same subcarrier, and thus a joint decoding technique is required. Since the convolutional encoding is adopted, an efficient sequence detection should be considered in the decoding scheme. In this section, we design a two-layer Viterbi-based algorithm for the joint TCM-based NOMA decoding in which the soft-decision based MLSD technique is utilized.
\subsection{Criterion for Joint Decoding}
We assume that $\eta$ data sequences are sent into each TCM encoder sequentially, i.e., one sequence per time unit. Denote $m$ as the number of time units required for the encoder registers to be cleared. The set of coded sequences is $\left( {{\textbf{c}}^{\left( k \right)}\left[ 1 \right],{\textbf{c}}^{\left( k \right)}\left[ 2 \right],\cdots,{\textbf{c}}^{\left( k \right)}\left[ {\eta} + m \right]} \right)$. In the traditional ML detection, the decoder should produce a set of estimated ${{\hat {\textbf{c}}}^{\left( k \right)}}$ of ${{\textbf{c}}^{\left( k \right)}}$ given the received sequence such that $\sum\limits_{t = 1}^{{\eta} + m} {\log P\left( {{y^{\left( k \right)}\left[t \right]}|{\textbf{c}}^{\left( k \right)}\left[t \right]} \right)}$ can be maximized.

However, due to the multiplexing nature of NOMA, the input data sequences of different encoders overlap with each other, which can be utilized for joint decoding. For example, given the mapping matrix in $\left( \ref{mapping_matrix} \right)$, the data streams of users 1, 2, and 3 are coded and transmitted over subcarrier 1, and that of users 1, 4, and 5 over subcarrier 2. A successful decoding scheme requires that the decoded data sequence of the same user obtained from different subcarriers in the one time unit should be the same. In other words, the data sequence of user 1 decoded from the first two subcarriers, i.e., ${\textbf{\emph{b}}}_1^{\left( 1 \right)}\left[t \right]$ and ${\textbf{\emph{b}}}_1^{\left( 2 \right)}\left[t \right]$, should be identical and the same is true with other users. Mathematically, we have
\begin{equation} \label{cross_check}
{\textbf{\emph{b}}}_j^{\left( k \right)}\left[t \right] = {\textbf{\emph{b}}}_j^{\left( {k'} \right)}\left[t \right], \mbox{if}\ j = {{\mathcal{J}}^{\left( k \right)}} \cap {{\mathcal{J}}^{\left( {k'} \right)}}, 1 \le k,k' \le K.
\end{equation}


Based on the above idea, we consider to adopt the cross check in the ML detection for joint NOMA decoding. Given the encoder state of time unit $\left( t - 1 \right)$ and the input data sequence of time unit $t$, a TCM encoder only produce one out of ${2^{q{d_f} + 1}}$ possible code sequences as the output. To depict this, we introduce a binary probability variable $P\left( {{\textbf{\emph{c}}}|{\textbf{\textbf{s}}}^{\left( k \right)}\left[t - 1\right],{\textbf{\emph{b}}}^{\left( k \right)}\left[t\right]} \right)$ in which ${\textbf{\emph{s}}}^{\left( k \right)}\left[t - 1\right]$ is the encoder state in time unit $\left( t - 1 \right)$. Given ${\textbf{\emph{s}}}^{\left( k \right)}\left[t - 1\right]$ and ${\textbf{\emph{b}}}^{\left( k \right)}\left[t\right]$, the probability of obtaining ${\textbf{\emph{c}}}$ over subcarrier $k$ is
\begin{equation} \label{probability_variable}
P\left( {{\textbf{\emph{c}}}|{\textbf{\emph{s}}}^{\left( k \right)}\left[t\!-\!1 \right],{\textbf{\emph{b}}}^{\left( k \right)}\left[t \right]} \right) =
\begin{cases}
1,  & \mbox{\shortstack[l]{if ${\textbf{\emph{c}}}$ is the output of\\ the encoder in unit $t$}}, \\
0,  & \mbox{otherwise}.
\end{cases}
\end{equation}
By combining condition $\left( \ref{cross_check} \right)$ and definition $\left( \ref{probability_variable} \right)$, we present the following proposition.

\textbf{Proposition 2:} Given the encoder states $\left\{ {{{\emph{\textbf{s}}}^{\left( k \right)}}\left[ t - 1 \right]|1 \le k \le K} \right\}$ in time unit $t - 1$, if $K$ code sequences in time unit $t$ are correctly estimated, then the following condition is satisfied:
\begin{equation} \label{probability_mul}
\prod\limits_{k = 1}^K {P\left( {{\hat {\emph{\textbf{c}}}^{\left( k \right)}}\left[ t \right]|{{\textbf{\emph{s}}}^{\left( k \right)}}\left[ t - 1 \right],{{\emph{\textbf{b}}}^{\left( k \right)}}\left[ t \right]} \right)}  = 1, \forall 1 \le t \le {\eta}+m,
\end{equation}
in which the kit of data sequences ${\emph{\textbf{B}}}\left[ t \right] = \left\{ {{{\emph{\textbf{b}}}^{\left( k \right)}}\left[ t \right]|1 \le k \le K} \right\}$ satisfies condition $\left( \ref{cross_check} \right)$.

For convenience, we denote a kit of code sequences as $\hat {\emph{\textbf{C}}}\left[ t \right] = \left\{ {{\hat {\emph{\textbf{c}}}^{\left( k \right)}}\left[ t \right]|1 \le k \le K} \right\}$. Since there are $K$ encoders, we define the super encoder state in time unit $t$ as the combination of encoder states of all $K$ encoders, i.e., ${\emph{\textbf{S}}}[t] = \left\{ {{{\textbf{\emph{s}}}^{\left( 1 \right)}}\left[ t \right],{{\textbf{\emph{s}}}^{\left( 2 \right)}}\left[ t \right],...,{{\textbf{\emph{s}}}^{\left( K \right)}}\left[ t \right]} \right\}$. Proposition 2 provides a necessary condition, and thus, there may exist multiple \emph{qualified} kits of code sequences in each time unit $t$. For any possible super encoder state ${\emph{\textbf{S}}}[t - 1]$, the set of qualified kits of coded sequences satisfying Proposition 2 in time unit $t$ is contained in ${\mathcal{C}}\left[ t \right]$.

The criterion for joint NOMA decoding can then be described as below. The decoder is required to estimate ${{\hat {\textbf{c}}}^{\left( k \right)}}$ given the received sequence $\textbf{y}$ by optimizing the following problem:
\begin{subequations} \label{ML_decoding}
\begin{align}
 \mathop {\max }\limits_{\hat{\emph{\textbf{C}}}\left[ t \right]} & \sum\limits_{t = 1}^{\eta + m} {\sum\limits_{k = 1}^K {\log P\left( {y^{\left( k \right)}\left[ t\right] |{{\hat {\emph{\textbf{c}}}}^{\left( k \right)}}\left[ t \right]} \right)} } \label{ML_1}\\
\textbf{\emph{s.t.}}\ & \hat{\emph{\textbf{C}}}\left[ t \right] \in {\mathcal{C}}\left[ t \right],\label{ML_2}
\end{align}
\end{subequations}
in which ${\sum\nolimits_{k = 1}^K {\log P\left( {y^{\left( k \right)}\left[ t\right] |{{\hat {\emph{\textbf{c}}}}^{\left( k \right)}}\left[ t \right]} \right)} }$ is called the branch metrics.

\subsection{Joint Decoding Scheme Design}
To solve the ML detection problem in $\left( \ref{ML_decoding} \right)$, we first present an optimal solution extended from the Viterbi algorithm~\cite{JP-1989}. To obtain a more practical solution with tolerable complexity, we then design a suboptimal two-layer Viterbi-based decoding algorithm.

\subsubsection{Optimal Maximum Likelihood Decoding Scheme}
For each encoder, we assume that the number of registers is $V$, and thus there are $2^V$ encoder states. Since there are $K$ encoders, we have $2^{KV}$ super encoder states in total. A super encoder state diagram can then be constructed based on the $2^{KV}$ super encoder states. Considering the cross check requirement of the NOMA decoder, we assume that one super encoder state can only transfer to another if the input data sequences and output coded sequences, i.e., ${\emph{\textbf{B}}}\left[ t \right]$ and ${\emph{\textbf{C}}}\left[ t \right]$, satisfy Proposition 2.

Given the super encoder state diagram, the soft-decision based Viterbi algorithm~\cite{JP-1989} can then be performed. For each super encoder state, we aim to update the survivor path in each time unit. A survivor path in time unit $t$ refers to a record consisting of a series of super encoder states ${{\emph{\textbf{S}}}\left[ t' \right]}$, $1 \le t' \le t$ and the branch between every two sequent states ${{\emph{\textbf{S}}}\left[ t'-1 \right]}$ and ${{\emph{\textbf{S}}}\left[ t' \right]}$. A \emph{branch} is denoted by the estimated input data sequences and output code sequences, i.e., ${\emph{\textbf{B}}}\left[ t' \right]$ and ${\emph{\textbf{C}}}\left[ t' \right]$. Mathematically, \emph{a survivor path} with respect to ${\emph{\textbf{S}}}\left[ t \right]$ can be given by
\begin{equation} \label{survivor_path}
\begin{split}
&{\emph{\textbf{S}}}\left[ 0 \right]\xrightarrow{{\emph{\textbf{B}}}\left[ 1 \right], {\emph{\textbf{C}}}\left[ 1 \right]} {\emph{\textbf{S}}}\left[ 1 \right] \to  \cdots  \to {\emph{\textbf{S}}}\left[ {t'} \right] \xrightarrow{{\emph{\textbf{B}}}\left[ t'+1 \right], {\emph{\textbf{C}}}\left[ t'+1 \right]} \\
& {\emph{\textbf{S}}}\left[ {t' + 1} \right] \to  \cdots  \to {\emph{\textbf{S}}}\left[ t \right].
\end{split}
\end{equation}

To update the survivor path for each ${{\emph{\textbf{S}}}\left[ t \right]}$, we utilize the branch metrics in $\left( \ref{ML_1} \right)$ to select a state ${{\emph{\textbf{S}}}\left[ t - 1 \right]}$ as well as the branch bridging ${{\emph{\textbf{S}}}\left[ t - 1 \right]}$ and ${{\emph{\textbf{S}}}\left[ t \right]}$. Specifically, the branch metrics in each time unit $t$ can be depicted by the total Euclidean distances between $K$ received signals and the signal points mapped to the candidate code sequences, i.e., $\sum\limits_{k = 1}^K {d\left( {y^{\left( k \right)}\left[ t\right] ,g\left( {{{\emph{\textbf{c}}}^{\left( k \right)}}\left[ t \right]} \right)} \right)}$, in which $g\left(  \cdot  \right)$ maps a coded sequence to the output signal point in the constellation.

In the last time unit ${\eta}+m$, we select the shortest survivor path terminated with an all-zero super encoder state as the final path. The optimality of this cross-check based soft-decision Viterbi algorithm can be guaranteed since the final survivor path is always the ML path (Theorem 12.1 in~\cite{SD-2004}).

\subsubsection{Suboptimal Two-layer Viterbi-based Joint Decoding Scheme}
In the optimal decoding algorithm, for each time unit the decoder is required to traverse all $2^{KV}$ super encoder states and possible branches to update the survivor path for each super encoder state, leading to a prohibitively high computation complexity. To reduce the complexity, we consider only checking the most possible super encoder states and branches to update no more than $\lambda$ survivor paths in each time unit in the suboptimal decoding scheme.


%

\textbf{a) Inner Layer Soft-decision Viterbi Operation:}
For convenience, we denote the set of stored super encoder states corresponding to the survivor paths in time unit $\left(t - 1\right)$ as ${\mathcal{S}}\left[ {t - 1} \right] = \left\{ {{{\emph{\textbf{S}}}_1}\left[ {t - 1} \right],{{\emph{\textbf{S}}}_2}\left[ {t - 1} \right],...,{{\emph{\textbf{S}}}_\lambda }\left[ {t - 1} \right]} \right\}$. For each ${{\textbf{\emph{S}}}_{i}}[t - 1]$, we first illustrate how to select the candidate branches based on the received signals over $K$ subcarriers in time unit $t$. The inner layer operation is performed over each subcarrier $k$ separately.

For a received signal point ${y^{\left( k \right)}\left[ t\right] }$, most candidate output signal points lie in the neighborhood of the received signal point since the noise power obeys the Gaussian distribution with a given variance ${\sigma ^2}$. Therefore, instead of recoding all the signal points corresponding to potential branches, we only record those lying in the neighborhood\footnote{The neighborhood of the received signal point can be defined by a circle whose center is this signal point.}.

To be specific, for each subcarrier $k$, given the current encoder state ${{{\textbf{\emph{s}}}_i^{\left( k \right)}}\left[ t - 1 \right]}$, the set of all possible input data sequences is denoted as $\cal{B}$ and $\left| {\mathcal{B}} \right| = {2^{q{d_f}}}$. The set of all possible output coded sequences can be denoted as ${\mathcal{C}}_{poss, i}^{\left( k \right)}\left[ t \right]$ such that each ${{\textbf{\emph{c}}}^{\left( k \right)}}\left[ t \right] \in {\mathcal{C}}_{poss, i}^{\left( k \right)}\left[ t \right]$ satisfies that
\begin{equation} \label{poss_defination}
P\left( {{{\textbf{\emph{c}}}^{\left( k \right)}}\left[ t \right]|{{\textbf{\emph{s}}}_i^{\left( k \right)}}\left[ t - 1 \right],{{\textbf{\emph{b}}}^{\left( k \right)}}} \right) = 1, \forall {{\textbf{\emph{b}}}^{\left( k \right)}} \in {\cal B}.
\end{equation}
The candidate output signal points over subcarrier $k$ can then be obtained by
\begin{equation} \label{inner_candidate}
\begin{split}
&{\mathcal{A}}^{\left( k \right)}_{cand, i}\left[ t \right] = \left\{ {A_l} \in {A_k}|d\left( {y^{\left( k \right)}\left[ t\right] ,{A_l}} \right) \le a{\sigma ^2},{A_l} = \right. \\
&\left. g\left( {{{\textbf{\emph{c}}}^{\left( k \right)}}\left[ t \right]} \right),{{\textbf{\emph{c}}}^{\left( k \right)}}\left[ t \right] \in {\mathcal{C}}_{poss, i}^{\left( k \right)}\left[ t \right] \right\},
\end{split}
\end{equation}
where $d\left( {y_t^{\left( k \right)},{A_l}} \right)$ is the Euclidean distance between ${y_t^{\left( k \right)}}$ and the signal point $A_l$, $a$ is the radius parameter. Based on the definitions of ${\mathcal{A}}^{\left( k \right)}_{cand, i}\left[ t \right]$ and ${\mathcal{C}}_{poss, i}^{\left( k \right)}\left[ t \right]$, we observe that each $A_l \in {\mathcal{A}}^{\left( k \right)}_{cand, i}\left[ t \right]$ corresponds to an output coded sequence and an input data sequence. For convenience, we denote them as members in the set of candidate coded sequences ${\cal C}_{cand, i}^{\left( k \right)}\left[ t \right]$ and the set of candidate data sequences ${\cal B}_{cand, i}^{\left( k \right)}\left[ t \right]$, respectively.


\textbf{b) Outer Layer Soft-decision Viterbi Operation:} In time unit $t$, given a stored super encoder state ${\emph{\textbf{S}}}_i[t - 1]$ and the outcome from the inner layer operation, we perform the cross check to obtain the survivor paths for time unit $t$.

For each super encoder state ${{\emph{\textbf{S}}}_i}\left[ {t - 1} \right]$, multiple candidate branches are generated based on ${\cal C}_{cand, i}^{\left( k \right)}\left[ t \right]$ and ${\cal B}_{cand, i}^{\left( k \right)}\left[ t \right]$. Specifically, a candidate branch $\left( {{{\emph{\textbf{C}}}_{cand,i}}\left[ t \right],{{\emph{\textbf{B}}}_{cand,i}}\left[ t \right]} \right)$ can be constructed by
\begin{equation} \label{outer_candidate}
\begin{array}{ll}
{\emph{\textbf{C}}}_{cand, i}\left[ t \right] = \left\{ {{{\emph{\textbf{c}}}^{\left( k \right)}}\left[ t \right]|{{\emph{\textbf{c}}}^{\left( k \right)}}\left[ t \right] \in {\mathcal{C}}_{cand,i}^{\left( k \right)}\left[ t \right],1 \le k \le K} \right\} \\
{\emph{\textbf{B}}}_{cand, i}\left[ t \right] = \left\{ {{{\emph{\textbf{b}}}^{\left( k \right)}}\left[ t \right]|{{\emph{\textbf{b}}}^{\left( k \right)}}\left[ t \right] \in {\mathcal{B}}_{cand,i}^{\left( k \right)}\left[ t \right],1 \le k \le K} \right\},
\end{array}
\end{equation}
in which $P\left( {{{\textbf{\emph{c}}}^{\left( k \right)}}\left[ t \right]|{{\textbf{\emph{s}}}_i^{\left( k \right)}}\left[ t - 1 \right],{{\textbf{\emph{b}}}^{\left( k \right)}}} \right) = 1$ for $1 \le k \le K$.

The set ${\cal{C}}\left[ t \right]$ (defined in Section III.A) can then be constructed by performing the cross check on the candidate branches. If a candidate branch $\left( {{{\emph{\textbf{C}}}_{cand,i}}\left[ t \right],{{\emph{\textbf{B}}}_{cand,i}}\left[ t \right]} \right)$ satisfies Proposition 2, we refer to it as a \emph{qualified branch}. For the qualified branch, we then add ${{\emph{\textbf{C}}}_{cand,i}}\left[ t \right]$ into the set ${\cal{C}}\left[ t \right]$ and record the corresponding super encoder state ${{\emph{\textbf{S}}}_i}\left[ t - 1\right]$.

Given the set ${\cal{C}}\left[ t \right]$, we then select at most $\lambda$ survivor paths for time unit $t$ based on the total path lengths. For each ${{\emph{\textbf{C}}}_{cand,i}}\left[ t \right] \in {\cal{C}}\left[ t \right]$, we can obtain the super encoder state ${{\emph{\textbf{S}}}}\left[ t \right]$ based on the corresponding input ${\emph{\textbf{B}}}_{cand, i}\left[ t \right]$ and the previous state ${{\emph{\textbf{S}}}_i}\left[ t - 1\right]$. The path length with respect to ${{\emph{\textbf{S}}}}\left[ t \right]$ is then calculated as
\begin{equation} \label{path_length}
\setlength{\abovedisplayskip}{2.5pt}
\setlength{\belowdisplayskip}{2.5pt}
\begin{split}
PLength\left( {{\emph{\textbf{S}}}\left[ t \right]} \right) & = PLength\left( {{\emph{\textbf{S}}}_i\left[ {t - 1} \right]} \right) \\
& + \sum\nolimits_{k = 1}^K {d\left( {y^{\left( k \right)}\left[ t\right] ,g\left( {{{\emph{\textbf{c}}}^{\left( k \right)}}\left[ t \right]} \right)} \right)},
\end{split}
\end{equation}
in which ${{{\emph{\textbf{c}}}^{\left( k \right)}}\left[ t \right]} \in {{\emph{\textbf{C}}}_{cand,i}}\left[ t \right]$ and $PLength\left( {{\emph{\textbf{S}}}_i\left[ {t - 1} \right]} \right)$ is the history total path length with respect to ${{\emph{\textbf{S}}}_i\left[ {t - 1} \right]}$.

Based on the above equation, we select up to $\lambda$ code sequences from ${\cal{C}}\left[ t \right]$ and the corresponding survivor paths with respect to various super encoder states ${{\emph{\textbf{S}}}\left[ t \right]}$.

\textbf{c) Overall Suboptimal Decoding Scheme:} The whole algorithm is presented in detail in Algorithm 2. In each time unit $t$, we aim to update the survivor paths based on the stored paths in time unit $t - 1$ and the received signals in time unit $t$. Specifically, for each subcarrier $k$ and each stored encoder state ${{\textbf{\emph{s}}}_{i}^{\left(k\right)}}[t - 1]$ $\left( 1 \le i \le \lambda \right)$, we first perform an inner layer soft-decision operation (line 5-9). The sets of candidate output coded sequences and input bit sequences are selected and temporally stored given the received signal $y^{\left( k \right)}\left[ t\right] $. We then perform the outer layer soft-decision Viterbi operation (line 10-20) in which the set of qualified branches satisfying Proposition 2 is selected based on the candidate coded sequences and bit sequences. By sorting the path lengths with respect to these selected branches, we update up to $\lambda$ survivor paths and super encoder states in time unit $t$ with the shortest path lengths (line 19-20). After $\eta + m$ time units, the shortest survivor path terminating at the all-zero super encoder state is selected as the final path (line 21-22). Different from the traditional Viterbi algorithm, the decoded bit sequence is already stored in each time unit, and thus, it can be directly obtained from the final path.

\begin{algorithm}[!h]
	\caption{Two-layer Soft-decision Viterbi Algorithm for Joint NOMA Decoding}\label{Alg_2}
	
	\hspace*{-0.02in} {\bf Input:} 
	$K$ received signal sequences ${{\emph{\textbf{y}}}^{\left( 1 \right)}},{{\emph{\textbf{y}}}^{\left( 2 \right)}},\cdots,{{\emph{\textbf{y}}}^{\left( K \right)}}$ of length $\eta + m$, the TCM encoder state diagram, NOMA mapping matrix $\emph{\textbf{F}}$, noise variance ${\sigma ^2}$, channel coefficients\\
	\hspace*{-0.02in} {\bf Output:} 

	$J$ decoded bit sequences for $J$ users

	\begin{algorithmic}[1]
	\State \textbf{Initialization:}
	\State Set ${{\emph{\textbf{S}}}}\left[ 0\right]$ as the all-zero super encoder state
	\State Set ${\cal{S}}\left[ 0 \right] = \left\{ {{\emph{\textbf{S}}}\left[ 0 \right]} \right\}$
	\For{$t = 1$ to $\eta + m$}
	\State Set ${\cal{C}}\left[ t \right] = \emptyset$
	\For {$i = 1$ to $\left| {{\cal{S}}\left[ {t - 1} \right]} \right|$}
	\For {$k = 1$ to $K$}
	\State Construct the set of candidate output signal points ${\mathcal{A}}^{\left( k \right)}_{cand, i}\left[ t \right]$ according to $\left(\ref{inner_candidate}\right)$
	\State Obtain the candidate sets ${\cal C}_{cand, i}^{\left( k \right)}\left[ t \right]$ and ${\cal B}_{cand, i}^{\left( k \right)}\left[ t \right]$ based on  $\left(\ref{poss_defination}\right)$ and $\left(\ref{inner_candidate}\right)$
	\EndFor
	\Repeat
	\State Construct the candidate branch $\left( {{{\emph{\textbf{C}}}_{cand,i}}\left[ t \right],{{\emph{\textbf{B}}}_{cand,i}}\left[ t \right]} \right)$ from ${\cal C}_{cand, i}^{\left( k \right)}\left[ t \right]$ and ${\cal B}_{cand, i}^{\left( k \right)}\left[ t \right]$ according to  $\left(\ref{outer_candidate}\right)$
	\If {the candidate branch never shows up and satisfies Proposition 2}
	\State Add ${{\emph{\textbf{C}}}_{cand,i}}\left[ t \right]$ to ${\cal{C}}\left[ t \right]$
	\State Record ${{\emph{\textbf{B}}}_{cand,i}}\left[ t \right]$ and ${{{\emph{\textbf{S}}}_i}\left[ {t - 1} \right]}$
	\EndIf
	\Until {there is no new candidate branch any more}
	\EndFor
	\For {each ${{\emph{\textbf{C}}}}\left[ t \right] \in {\cal{C}}\left[ t \right]$}
	\State Obtain ${{{\emph{\textbf{S}}}}\left[ t \right]}$ based on the recorded ${{\emph{\textbf{B}}}_{cand}}\left[ t \right]$ and ${{{\emph{\textbf{S}}}}\left[ {t - 1} \right]}$
	\State Calculate the path length $PLength\left( {{\emph{\textbf{S}}}\left[ t \right]} \right)$ according to $\left( \ref{path_length} \right)$
	\EndFor
	\State Sort the obtained path lengths and update the first $\lambda$ paths with the shortest lengths
	\State as survivor paths (shown in $\left( \ref{survivor_path} \right)$)
	\EndFor
	\State Select the shortest survivor path in time unit $\eta + m$ which terminates at the all-zero super
	\State encoder state as the final path
	\State \Return the decoded bit sequences corresponding to the final path
	\end{algorithmic}
\end{algorithm}

\textbf{d) Computational Complexity Analysis:} The computational complexity of Algorithm \ref{Alg_2} in each time unit $t$ comes from three parts:
\begin{itemize}
\item \emph{Inner layer operation (line 6-9):} For each super encoder state ${{\emph{\textbf{S}}}}_i\left[ t - 1 \right]$ and each subcarrier $k$, at most $2^{q{d_f}}$ bit sequences are checked to construct ${\cal C}_{cand, i}^{\left( k \right)}\left[ t \right]$ and ${\cal B}_{cand, i}^{\left( k \right)}\left[ t \right]$ based on  $\left(\ref{poss_defination}\right)$ and $\left(\ref{inner_candidate}\right)$. Since there are $K$ subcarriers and at most $\lambda$ stored super encoder states in time unit $t$, the complexity of the inner layer operation is $O\left( {\lambda K{2^{q{d_f}}}} \right)$. Note that in a practical NOMA scheme, we usually set $q{d_f} \le 6$ such that the complexity is tolerable.
\item \emph{Searching for qualified branches (line 10-15):} Suppose that the number of candidate bit sequences for each super encoder state ${{\emph{\textbf{S}}}}_i\left[ t - 1 \right]$ and each subcarrier $k$ is $M_i^{\left( k \right)}$ and $M_i^{\left( k \right)} \le {2^{q{d_f}}}$. For each ${{\emph{\textbf{S}}}}_i\left[ t - 1 \right]$, the number of candidate branches is $\prod\nolimits_{k = 1}^K {M_i^{\left( k \right)}}$. The process of checking whether a candidate branch satisfies Proposition 2 can be performed when constructing such a branch. Therefore, the complexity of constructing all qualified branches is $O\left( {\lambda \prod\nolimits_{k = 1}^K {M_i^{\left( k \right)}} } \right)$. In practice, the value of ${M_i^{\left( k \right)}}$ is directly influenced by the radius parameter $a$ and ${\sigma ^2}$ in $\left( \ref{inner_candidate} \right)$, as will be discussed in Section V.
\item \emph{Updating survivor paths (line 16-20):} Suppose that the number of qualified branches\footnote{The upper bound of $Q_i$ is closely related to the mapping matrix $\emph{\textbf{F}}$. For example, based on $\emph{\textbf{F}}$ in $\left( \ref{mapping_matrix} \right)$, we have ${Q_{\max }} = M_i^{\left( 1 \right)} \cdot \min \left\{ {{2^{q\left( {{d_f} - 1} \right)}},M_i^{\left( 2 \right)}} \right\} \cdot {2^q}$. The detailed proof is omitted due to the limitation of space.} given a super encoder state ${{\emph{\textbf{S}}}}_i\left[ t - 1 \right]$ is $Q_i$, i.e., $\left| {{\mathcal{C}}\left[ t \right]} \right| = \sum\nolimits_{i = 1}^\lambda  {{Q_i}}$. To update the survivor paths, we need to sort the qualified branches and the complexity can be given by $O\left( {{{\left( {\sum\nolimits_{i = 1}^\lambda  {{Q_i}} } \right)}^2}} \right)$.
\end{itemize}
The total complexity of Algorithm 2 can be obtained by adding up the above three parts. We will present detailed results in Fig.~\ref{complexity} in Section V.

\section{Simulation Results}
In this section, we consider a NOMA system as shown in Fig.~\ref{system_model}. For the convolutional encoder in AWGN channels, the coefficients of the parity-check sequences in octal form are selected based on the look-up table in~\cite{SRAGD-1990}. For simplicity, we only assume perfect channel state information at the receiver. The link-level performance of our proposed TCM-based NOMA scheme is compared with the traditional 4-point code-domain NOMA scheme in~\cite{HH-2013}. To evaluate the design of the MD-constellation for TCM-based NOMA, we also propose a lattice constellation TCM (LC-TCM) based NOMA scheme as comparison, in which the constellation ${\cal{A}}^{\left( k\right) }$ constructed for each subcarrier $k$ is the same as the traditional 2D lattice constellation~\cite{GL-1989} of size ${2^{q{d_f} + 1}}$. Major simulation parameters are listed in Table~\ref{table_simulation}.

\begin{table}
	\begin{center}
		\caption{Major Simulation Parameters}
		\label{table_simulation}
		\begin{tabular}{|p{55mm}|l|}
			\hline
			\bf{Parameter} & \bf{Value}\\
			\hline The number of users sharing the same subcarrier $d_f$ &3\\
			\hline The number of non-zero elements in a sparse codeword $N$ &2\\
			\hline The number of bits to be coded for a user in each time unit $q$ &2\\
			\hline Convolutional encoder rate $r/{\left( r+1\right) }$ &3/4\\
			\hline MD mother constellation &$d_f \times $16QAM\\
			\hline Frame length & 1000 bits\\
			\hline Spectral efficiency & 1.5 or 3 bis/tone\\
			\hline radius parameter $a$ & 4 $\sim$ 6 \\
			\hline Maximum number of survivor paths in each time unit $\lambda$ & 5 $\sim$ 35\\
			\hline Number of register states $V$ & 4\\
            \hline Interleaver size & 32$\times$16 (in symbol)\\
            \hline Maximum Doppler shift in Rayleigh channels & 50 Hz\\
            \hline Sampling frequency for Rayleigh channels & 1/1800 s\\
			\hline
		\end{tabular}
	\end{center}
\end{table}

\begin{figure}[!ht]
	\centering
	\includegraphics[width=3.6in]{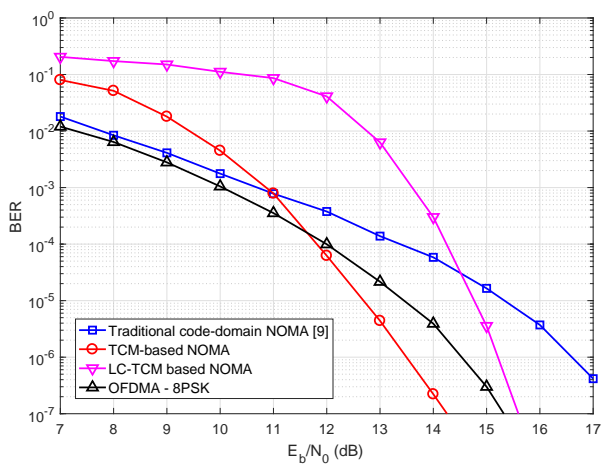}
	\caption{BER v.s. $E_b/N_0$ over AWGN channels without channel coding.} \label{AWGN-BER}
\end{figure}

\begin{figure}[!ht]
	\centering
	\includegraphics[width=3.6in]{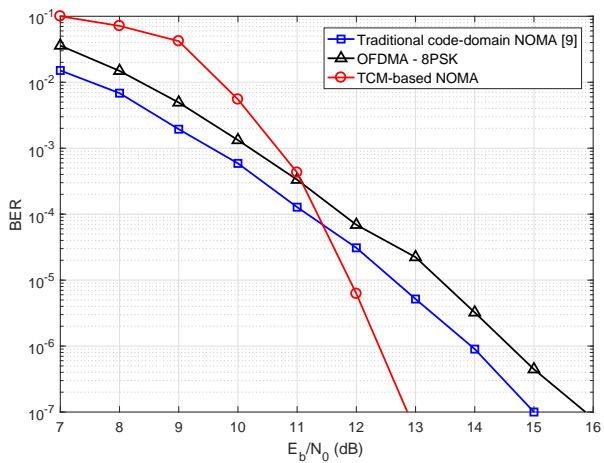}
	\caption{BER v.s. $E_b/N_0$ over AWGN channels with convolutional channel coding of rate 1/2.} \label{Coding-AWGN-BER}
\end{figure}

\begin{figure}[!ht]
	\centering
	\includegraphics[width=3.5in]{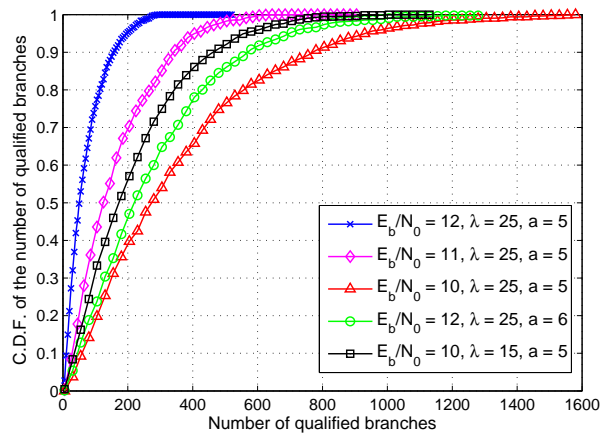}
	\caption{C.D.F. of the number of qualified branches in Algorithm 2.} \label{complexity}
\end{figure}

\begin{figure}[!ht]
	\centering
	\includegraphics[width=3.5in]{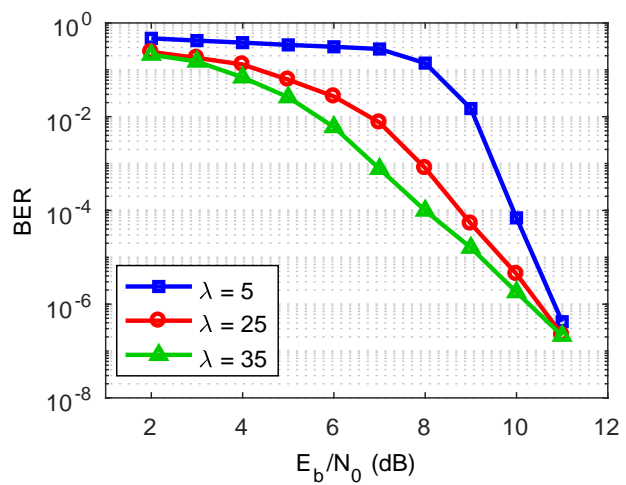}
	\caption{BER performance v.s. $E_b/N_0$ with different number of survivor paths.} \label{setSize}
\end{figure}

\begin{figure}[!ht]
	\centering
	\includegraphics[width=3.6in]{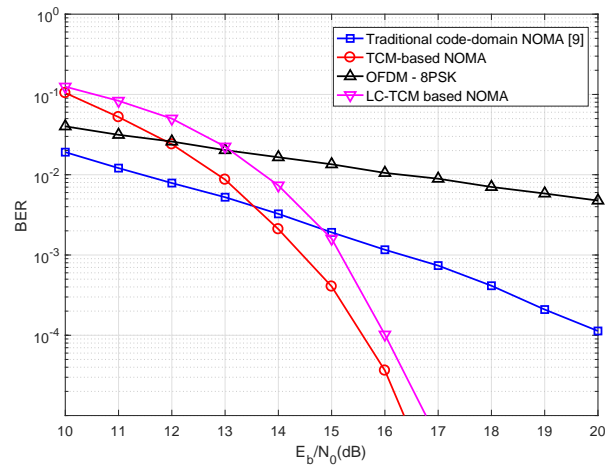}
	\caption{BER v.s. $E_b/N_0$ over Rayleigh fading channels.} \label{Ray-BER}
\end{figure}

Fig.~\ref{AWGN-BER} and Fig.~\ref{Coding-AWGN-BER} illustrate the BER performance versus the $E_b/N_0$ over an AWGN channel with and without channel coding of rate 1/2, respectively. We set $\lambda = 25$ and $a = 5$, and the spectrum efficiency are 3 bits/tone and 1.5 bits/tone, respectively. In both figures, the BER performance of the TCM-based NOMA outperforms that of the traditional code-domain NOMA scheme with high $E_b/N_0$ and the gap between them increases with $E_b/N_0$. When $E_b/N_0$ is small, the error correcting capacity of the TCM system is restricted due to the high noise level, which is very common in the coded systems. This leads to a worse performance than the traditional scheme. However, as $E_b/N_0$ increases, the influence of bursty channel errors on the TCM-based NOMA becomes smaller, and thus, the performance gain obtained from the joint coded modulation grows. Fig.~\ref{AWGN-BER} also shows that the TCM-based NOMA outperforms the LC-TCM based NOMA scheme. The lattice constellation applied in the LC-TCM based scheme fails to optimize the positions of multiple MD points, leading to a small MFSE distance of the system and bad performance compared to the proposed scheme. This implies that the construction of the MD-mother constellation is of vital importance for joint coded modulation.

Specifically, In Fig.~\ref{AWGN-BER} we observe that without channel coding, the BER performance of the traditional code-domain NOMA is worse than that of the OFDMA scheme due to the existence of multi-user interference. Benefited from the joint codeword design of different users, our proposed TCM-based NOMA can achieve a better BER performance than both the OFDMA scheme and the traditional NOMA scheme when $E_b/N_0$ is over 11 dB. This implies that TCM-based NOMA provides a method to compensate for the degraded performance of traditional NOMA when the channel coding is not considered. In Fig.~\ref{Coding-AWGN-BER}, the traditional NOMA scheme outperforms the OFDMA scheme when channel coding is adopted, which is in consistent with existing works on NOMA.

To evaluate the computational complexity of our proposed decoding scheme, we adopt the number of qualified branches for each time unit $t$ as the metrics, i.e., $\left| {{\mathcal{C}}\left[ t \right]} \right| = \sum\nolimits_{i = 1}^\lambda  {{Q_i}}$, mentioned in Section IV.B.2.d. Fig.~\ref{complexity} shows the cumulative distribution function (C.D.F.) of $\left| {{\mathcal{C}}\left[ t \right]} \right|$, $\Pr \left( {\left| {{\cal C}\left[ t \right]} \right| \le \tilde c} \right)$, versus ${\tilde c}$ for different $E_b/N_0$ and decoding parameters. For a fixed number of limited survivor paths $\lambda$ and radius parameter $a$, the complexity grows as $E_b/N_0$ decreases. A smaller $E_b/N_0$ implies a higher noise level ${\sigma ^{\rm{2}}}$, and thus the number of candidate output signal points grows according to $\left(\ref{inner_candidate}\right)$, leading to more candidate branches and higher complexity. This also reveals the reason that the complexity increases with the radius parameter $a$ when $E_b/N_0$ and $\lambda$ are fixed. Besides, Fig.~\ref{complexity} also shows that the complexity of our proposed decoding scheme decreases with $\lambda$ becoming smaller.

Fig.~\ref{setSize} shows the BER performance of the TCM-based NOMA scheme versus the $E_b/N_0$ over an AWGN channel for different numbers of survivor paths $\lambda$. As $\lambda$ increases, the BER performance improves since the probability that a correct path can be found in the survivor paths is larger. Note that when $E_b/N_0$ is too small or high enough, the increase of $\lambda$ does not have much influence on the BER performance. When $E_b/N_0$ is small, it is still hard for the decoder to select the right path from a large number of survivor paths due to interference caused by high noise level. When $E_b/N_0$ is large enough, the correct path is usually the shortest survivor path for each time unit, and thus, it is not necessary to increase $\lambda$ any more. Since both the BER performance and the complexity grow with $\lambda$, a trade-off can be achieved between these two metrics.

Fig.~\ref{Ray-BER} presents the BER performance versus the $E_b/N_0$ over a Rayleigh channel with $\lambda = 25$ and $a = 6$. Compared to the case with an AWGN channel, the encoder state diagram is modified as mentioned in Section III.B.3 and the interleaving across the time-frequency resources is performed. Since the number of branches leaving from one state is larger than that of the registers in the convolutional encoder, the shortest error path length is limited. Therefore, when $E_b/N_0$ is small, the bursty channel errors of the fading channels may be beyond the scope of the system's error control capacity. As $E_b/N_0$ grows, the performance gain brought by joint encoding exceeds the negative effects caused by the noise level. Thus, our proposed scheme significantly outperforms the traditional NOMA scheme when $E_b/N_0$ is larger than 13dB. Fig.~\ref{Ray-BER} also shows that 1) the carefully designed MD constellation in the TCM-based NOMA scheme brings better BER performance than the lattice constellation in the LC-TCM scheme; 2) the traditional code-domain NOMA outperforms the OFDMA scheme in Rayleigh fading environments.

\section{Conclusions}
In this paper, we have proposed a TCM-based NOMA scheme in which the codewords of multiple users are jointly designed based on the MD-TCM techniques, aiming at improving the coding gain of the system. For the encoder design, an MD constellation is constructed and the signal set is selected and labeled to maximize the MFSE distance of the system. To perform the joint decoding, the detection criteria for MLSD are formulated in which the non-orthogonal nature is utilized. A suboptimal two-layer soft-decision Viterbi decoding scheme is then proposed. Simulation results have showed that the TCM-based NOMA scheme performed better than the traditional code-domain NOMA scheme and OFDMA scheme in terms of the BER performance for both the AWGN channels and the Rayleigh fading channels with or without channel coding.

\begin{appendices}
\section{Proof of Proposition 1}
As shown in Algorithm 1, in each FPO iteration, a point $A_l \in {\cal{A}}_b^1$ is swapped with the farthest point of ${\cal{A}}_b^1$, i.e., $A_{r^*} \in {\cal{A}}_b^2$. Based on line 35-36 in Algorithm 1, after swapping the minimum distance between $A_{r^*}$ and ${\cal{A}}_b^1$ is larger than that between $A_l$ and ${\cal{A}}_b^1$, i.e.,
\begin{equation} \label{inequal1}
{d_{{r^*}}}\left( {{\cal{A}}_b^1 \cup \left\{ {{A_{{r^*}}}} \right\}\backslash {A_l}} \right) > {d_l}\left( {{\cal{A}}_b^1} \right),
\end{equation}
and the minimum distance between $A_l$ and ${\cal{A}}_b^2$ is no smaller than the MSSD of ${\cal{A}}_b^2$. Based on $\left( \ref{MSSD2} \right)$, we can infer that the MSSD of ${\cal{A}}_b^1$ is also increasing since the minimum distance of each point in ${\cal{A}}_b^1$ is increasing. According to $\left( \ref{average_minimum_distance} \right)$ and $\left( \ref{inequal1} \right)$, the average minimum distance of the updated subset ${\cal{A}}_b^1$, i.e., $\bar d\left( {{\cal{A}}_b^1} \right)$, is larger than that of the original subset ${\cal{A}}_b^1$ after one FPO iteration. Since there always exists an upper bound of $\bar d\left( {{\cal{A}}_b^1} \right)$, Phase 2 will stop within a limited number of iterations.

\section{Proof of Complexity of Algorithm 1}
For convenience, we assume that $\left| {{{\cal{A}}_b}} \right| = N$. In Phase 1, the complexity of the sorting process in line 2 is $O\left( {L\log L} \right)$ with $L = C_N^2$. The insertion of a point into the DT is $O\left(1\right)$, and thus, the complexity of the greedy algorithm (line 5-22) is $O\left( N \right)$. Therefore, the complexity of Phase 1 is $O\left( {{N^2}\log N} \right)$.

For each FPO iteration of Phase 2, the complexity of the DT-SEARCH operation is $O\left(g\right)$ in which $g$ is the number of a point's Delaunay neighbors. It is commonly assumed that $g = O\left(1\right)$ for well-distributed point sets~\cite{E-2005}. From line 25-37, the complexity is $O\left(N^2\right)$ in the worst case. By using a binary search tree to update the vertices in the DT, the complexity of one FPO iteration can be reduced to $O\left( {N\log N} \right)$~\cite{TDO-2011}.
\end{appendices}

\end{document}